# Exploring conformational energy landscape of glassy disaccharides by CPMAS 13C NMR and DFT/GIAO simulations. II. Enhanced molecular flexibility in amorphous trehalose.

Ronan LEFORT[1], Patrice BORDAT[2], Attilio CESARO[3], and Marc DESCAMPS[1]

[1]Laboratoire de Dynamique et Structure des Matériaux Moléculaires, P5, Université de Lille 1, Cité Scientifique, F-59655 Villeneuve d'Ascq Cedex, France

[2]Laboratoire de Chimie Théorique et de Physico-Chimie Moléculaire, UMR 5624 - IFR, LCS, 2, rue Jules Ferry, F-64000 Pau, France

[3]Laboratory of Physical and Macromolecular Chemistry, University of Trieste, Via Giorgieri 1, I-34127 Trieste, Italy

Abstract :

This paper deals with the comparative use of the chemical shift surfaces to simulate experimental $^{13}$C CPMAS data on amorphous solid state disaccharides, paying particular attention to $\alpha$-1-1 linkage of trehalose, to $\beta$-1,4 linkage between pyranose rings (lactose) and to linkage implying a furanose ring (sucrose). The combination of molecular mechanics with DFT/GIAO *ab-initio* methods provides reliable structural information on the conformational distribution in the glass. The results are interpreted in terms of an enhanced flexibility that trehalose experiences in amorphous solid state compared to the other sugars. An attempt to relate this property to the balance between intra- and inter-molecular hydrogen bonding network in the glass is presented.

PACS : 61.43.Bn, 61.18.-Fs , 76.60.Cq

**Introduction**

The general understanding of the local structure of inorganic amorphous states, like metallic or oxide glasses, has progressed through giant steps in the last decades, both on the experimental and theoretical fronts [1-3]. Due to the rather simple form of the elementary unit in those systems, the details of the local structure can often be very accurately described using a coordination shell image, and most of the information can be revealed through radial pair-distribution functions, or their reciprocal counterpart (structure factor) which is accessible through neutron or x-ray diffraction experiments. This approach rapidly reveals to be incomplete when molecular dissymmetry or specific and directional interactions like hydrogen bonding are present. Building a correct structural model of disordered hydrogen-bonded networks remains today a fascinating challenge. As the archetype system, studies on water are innumerable (see for example [4-8]). Other hydrogen-bonded glasses like protein networks, polyalcohols or saccharides have also recently raised considerable interest, since they meet the frontiers of biophysics, food, pharmacy, and materials science [9, 10]. The rich complexity of carbohydrate systems compared to model atomic glasses originates from their large number of intramolecular degrees of freedom. These coordinates necessarily affect the dynamic properties of the compound near its glass transition, often conferring a fragile character to the structural relaxation of the glass former [11] and also making the description of the glass structure a very complicated task. Only very recently, local investigation techniques like nuclear magnetic resonance (NMR) or numerical experiments by molecular dynamics (MD) have been proved to be very useful in [12-14].

The special case of disaccharides deserves a particular attention, for which a relatively small number of geometrical parameters are suspected to give a reasonable picture of their intramolecular energy landscape [13-18], conferring emphasis to the stereochemistry of these compounds. Within this family of carbohydrates, trehalose has focused considerable interest, mainly because of its exceptional biopreservation properties, with a mechanism that still remains an open question at the molecular level [19-22]. The same sugar has also been proposed as a model hydrogen-bond network well described by the axiomatic theory of ideally glassy network developed by Phillips and co-workers [9, 10]. Molecular modelling by empirical force-fields is a well adapted technique to explore the conformational properties of carbohydrates [13, 16, 17, 23-27]. For many species, the large number of parameters that define the molecular geometry give raise to complicated high dimensional energy maps. These maps often present several potential wells, that correspond to different preferred conformations, and the question whether the calculated conformations are actually populated or not in solution and in pure disordered phases is still a matter of debate [13, 14, 27, 28]. The comparison of these numerical simulations with experiment is therefore a very important and demanding step, where local techniques like NMR already prove to be determinant [12, 18, 29, 30].

The aim of the present work is to analyse and compare the conformational space that is actually accessible to three homologous sugars (trehalose, sucrose and lactose) in their respective pure amorphous phases. The general method relies on the comparison between the preferred molecular geometries predicted by classical mechanics via empirical force-fields, and the actual distribution of conformations that is observed by $^{13}$C solid state NMR on amorphous samples. The scientific background of the method and the simulation set-up have been discussed in detail elsewhere [18]. After a brief description of the experimental and computational conditions (section 2), we detail the results obtained by $^{13}$C CPMAS NMR experiments. We then report the results of the simulations, which are then discussed in relation to the balance between intra- and inter-molecular interactions, and their influence on the local structure of the hydrogen-bonded glassy network.

## Experiments and methods

**Samples.** Anhydrous α,α-trehalose (T$_\beta$), stable anhydrous α-lactose and sucrose crystalline powders were purchased (Sigma-Aldrich) and used without further purification. The amorphous samples were obtained by ball-milling 1g of crystalline sugar during 30 hours under dry nitrogen atmosphere in a planetary grinder PULVERISETTE 7 (Fritsch, inc.). A DSC thermogram was recorded after grinding in order to check the total character of the amorphization, and the glassy character of the obtained amorphous sample.

**NMR Experiments.** The $^{13}$C CPMAS experiments were carried out at 100.6 MHz on a Bruker AV400 solid-state NMR spectrometer. Linear amplitude modulation of the rf field during the contact pulse (typ. 2 ms), and tppm heteronuclear decoupling during acquisition were employed. Recycle delays ranging between 200 and 450 s were used. Rotation speed was set to 5 kHz. A standard digital filter was used for acquisition, and the spectra were obtained by simple Fourier transform of the induction decay, without data apodization.

**Molecular modelling.** Molecular modelling was achieved using the Hyperchem Pro 7 (Hypercube, Inc.) software package. Conformations were generated by relaxing all the degrees of freedom but constraining the two dihedral angles $\phi$ and $\psi$, as defined for trehalose, lactose and sucrose in Table 1. For fixed $\phi$ and $\psi$, all other conformational degrees of freedom were optimized using the molecular mechanics package provided in Hyperchem, including equivalent implementations of the AMBER and CHARMM empirical force fields. The parameters version used will be denoted in the text by BIO85 (Reiher, [31]). In order to take into account a mean-field contribution of the surrounding molecules in a condensed matter sample, a standard screening procedure of coulomb interaction was used, through an effective scaled dielectric constant. The 1-4 Scale Factors were equilibrated between Coulomb and Van der Waals interactions.

**Conformational energy maps.** All calculations were started from the molecular coordinates obtained by the single crystal structure determination on anhydrous trehalose ($T_\beta$) [32], lactose monohydrate ($\alpha L_{H_2O}$) [33] and sucrose [34]. Each of these crystal structures was relaxed by molecular mechanics in order to define the lowest energy molecular conformation (LMC) in the BIO85 force field. For each sugar, an adiabatic Ramachandran map $E(\phi,\psi)$ was calculated by classically mapping the whole $(\phi,\psi)$ space by 324 points separated by 20° steps [12, 30, 35]. For each $(\phi,\psi)$ point in the map, the $\phi$ and $\psi$ values were set starting from the LMC coordinates, and then were restrained by setting a high value of dihedral angle spring constants. All other degrees of freedom were then relaxed using the BIO85 force field parameters.

**Isotropic $^{13}C$ chemical shift maps.** NMR chemical shift calculations were carried out for each studied disaccharide on the previous 324 relaxed conformations generated by molecular mechanics. For each conformation, the isotropic magnetic shielding of each carbon of the sugar was evaluated using the GIAO method on a density functional theory (B3PW91) with the 3-21+G** basis, as implemented in the Gaussian 03 software (Gaussian, Inc.) [36]. The 324 single point calculations resulted for each sugar in two magnetic shielding maps, that were converted into chemical shift maps $\sigma(C_x,\phi,\psi)$ and $\sigma(C_y,\phi,\psi)$, where $C_x$ and $C_y$ represent the two carbons involved in the glycosidic bond of the disaccharide. The conversion from rough magnetic shielding to chemical shift comparable to the experiment was carried out by simple linear transformation $\sigma = A.S_{calc.} + B$, where the A and B coefficients are given in Table 2.

**Simulation of the CPMAS spectrum.** The scientific background leading to the formulation of the simulated $S(\delta)$ CPMAS spectrum was presented and discussed in detail elsewhere [18]. The numerical evaluation of $S(\delta)$ requires integration of constant chemical shift contours in the map $\sigma(C_x,\phi,\psi)$, using equation (1).

$$S(\delta) = \oint_{\sigma(\phi,\psi)=\delta} \frac{\chi(\phi,\psi)}{\left|\vec{\nabla}\sigma(\phi,\psi)\right|_{\sigma=\delta}} d\phi.d\psi \qquad (1)$$

In this equation, $\chi(\phi,\psi)$ represents the probability density of occupation of a given molecular conformation $(\phi,\psi)$ [18]. This function $\chi(\phi,\psi)$ thoroughly carries the structural information that can be extracted out of the modelling of the NMR spectrum. The actual resolution of this map is limited in the present work to 20° angular steps. In order to smooth the integration contour, cubic spline interpolation was used. The simulated CPMAS spectrum $S(\delta)$ was then evaluated by sampling 1024 different $\delta$ values, each one corresponding to one numerical integration over an interpolated $\sigma(C_x,\phi,\psi)=\delta$ contour, evaluated by standard Romberg method. The gradient term in Eq. (1) was calculated by finite difference approximation using $\Delta\phi=\Delta\psi=0.01°$ angular steps.

The last step of data processing is the convolution of $S(\delta)$ with a lorentzian function, in order to account for the homogeneous CPMAS linewidth. This was achieved by inverse Fourier Transform of $S(\delta)$ to a free induction decay $S(t)$, then multiplication of $S(t)$ by an exponential function and back Fourier Transform to $S(\delta)$. A typical line broadening imposed in this work was 1 ppm HWHM.

## $^{13}$C CPMAS Spectra

The Figure 1 compares the CPMAS spectra of crystalline and amorphous forms of three disaccharides. In all cases, the particular resonances assigned to the two carbons linked to the bridging oxygen are marked.

The spectral feature common to the three sugars in the amorphous state is the relevant broadening of all NMR signals. These distributions of isotropic chemical shifts reflect the underlying distributions of molecular conformations in the glass, which largely rely upon the changes in the glycosidic dihedral angles. For the glycosidic carbons, the observed chemical shift distributions spread over about 10 ppm for sucrose and lactose. For trehalose, it is highly asymmetric, tails downfield, and extends over more than 15 ppm. This particularity would suggest that the trehalose molecules experience enhanced flexibility in the glass compared to the other disaccharides, if the chemical shift maps are similar. Another remarkable point for the three sugars is that the NMR spectrum of the amorphous states does not completely envelope that of their respective crystalline states. This is especially noticeable for the glycosidic carbons of lactose and sucrose, for which the maxima of the chemical shift distributions do not coincide with the NMR lines of the crystal. For lactose, we observe an upfield shift of the NMR distributions, particularly important for the glucose residue $C'_4$ resonance. For sucrose, the observed shift is downfield. For trehalose, this effect is not pronounced, and the maximum of the distribution in the glass exactly corresponds to one of the $C_1/C'_1$ resonances of the $T_\beta$ crystal. These observations support the idea that the most probable molecular conformation in the glassy state is different, at least for lactose and sucrose, from the one observed in the crystalline form.

# Simulations

This section illustrates the steps leading to the simulations of the CPMAS spectra. First, the relaxed molecular conformations are generated by molecular mechanics, minimizing the corresponding potential energy surfaces $E(\phi,\psi)$, for each pair $(\phi,\psi)$. Then, for each conformations the corresponding $\sigma(C_x,\phi,\psi)$ maps for the glycosidic carbons are calculated.

## Energy and Chemical shift surfaces

Figure 2 presents the potential energy maps obtained by relaxing a single molecule (starting form the conformation of the molecule in the crystal) by CHARMM-type molecular mechanics with the BIO85 parameters. Rather that the global minimum, here the minimum energy conformation of the molecule constrained in each given $\phi,\psi$ pair is searched. Only the region of the map around the lowest energy basin is shown. It appears clearly from Table 3 that the preferred conformation of lactose and sucrose in the crystal is not located in the centre of the region with the lowest energy but differs significantly from the calculated one (although still in the low energy basin). It clearly means that the single molecule geometry is affected by the intermolecular interactions in the crystalline field, in the order lactose > sucrose > trehalose. At least qualitatively, this difference accounts for the shifts observed in Figure 1 between the spectra for amorphous and crystalline disaccharides. One may wonder whether the information provided by the comparison between NMR data for amorphous and crystalline forms and that obtained by the conformational energy maps provide any hints about the validity of the computational procedure.

Single molecule approach should be considered simplistic as it neglects the potential mean field generated by the random distribution of other sugar molecules. Undoubtedly, intermolecular interactions, which are relevant in the periodic crystalline field, may have less significance in a merely statistical distribution. Furthermore, slow dynamics averages may hamper the effectiveness of the calculation, while the difference in the internal conformational population may result less significant.

In the case of trehalose the calculated and crystallographic torsional angles defining the geometry of the glycosidic bond remain close together. These observations are consistent with the analysis of the experimental NMR spectra, and suggest that a structural description of amorphous forms of disaccharides based only on single molecule properties could be valid in a first order approximation.

Following this hypothesis, the simulation of the CPMAS spectra was carried out based on such a single molecule approach.

Figure **3** shows the chemical shift surfaces, calculated over the molecular conformations shown in the $E(\phi,\psi)$ maps. As a consequence of the molecular symmetry of trehalose, the equivalence $E(\phi,\psi)=E(\phi,\psi)^T$ and $\sigma(C_1,\phi,\psi)= \sigma(C'_1,\phi,\psi)^T$ are evident in Figures 3 and 4. This characteristic property has been profitably used to average the small fluctuations of the calculations.

**Results and discussion**

Figure 4 presents the simulated regions of the CPMAS spectra corresponding to the glycosidic carbons of $\alpha,\alpha$-trehalose, $\alpha$-lactose and sucrose. Simulations were carried out according to equation 1, and calculated over the contours of the chemical shift maps presented in Figure **3** by using the appropriate $\chi(\phi,\psi)$ functions. Two different forms of the $\chi(\phi,\psi)$ molecular conformational probability were tested. First, a Boltzmann partition on the $E(\phi,\psi)$ maps presented in Figure 2 has been assumed, see equation (2), by using the the temperature $T_B$ as an adjustable parameter for fitting:

$$\chi(\phi,\psi) \propto \exp\left[-\frac{E(\phi,\psi)}{R.T_B}\right] \quad (2)$$

A second procedure assumed a Gaussian partition of conformations, with maximum centred on $(\phi_g,\psi_g)$ and width parameters $\Delta\phi$ and $\Delta\psi$ defined by equation (3):

$$\chi(\phi,\psi) \propto \exp\left[-\frac{(\phi-\phi_g)^2}{2\Delta\phi^2}-\frac{(\psi-\psi_g)^2}{2\Delta\psi^2}\right] \quad (3)$$

where $\phi_g$, $\psi_g$, and $\Delta\phi = \Delta\psi$ were adjustable parameters for the simulations. This second procedure was included to formulate a set of "equivalent" energy basins that provides the quantitative comparison for flexibility.

As shown in Figure 4, the simulations with Boltzmann populations $\chi(\phi,\psi)$ of the $E(\phi,\psi)$ maps (Figure 2), *i.e.* with reference to a single molecule energy landscape model, poorly describe the experimental data, except in the case of trehalose. However, a very large temperature value had to be assumed in the equation (2) for this system. We shall return to this point in the discussion. It has been already showed [18] that the most important parameter in those calculations is rather the position of $\chi(\phi,\psi)$ than its very detailed shape. As a consequence, simulations seem to indicate that either the conformational populations resulting from molecular mechanics modelling or the chemical shift surfaces contain approximation. Given the procedure followed which assigns the highest accuracy to the calculation of the $\sigma(C_x,\phi,\psi)$ maps, attempts have been made to search for a more effective fit of the chemical shift distributions by adjusting the energy maps. As a matter of fact, although the obvious choice would have been a novel, but time consuming, calculation including the potential mean field generated by the random distribution of other sugar molecules, a satisfactory agreement between experiments and calculations were obtained by using a simple gaussian conformational occupation $\chi(\phi,\psi)$, slightly shifted apart from the minima of $E(\phi,\psi)$, as reported in Table 4. Although not necessarily realistic, this result once more reveals the great influence of the highest probability region on the $\sigma(C_x,\phi,\psi)$ map. With such artificial conformational distribution a good description of the experimental line shapes is obtained, except for the lactose $C'_4$ NMR line.

At this point, it is instructive to compare the angular features of the conformational states that provide a reasonable fit to the experimental NMR chemical shift data (compare Table 3 and Table 4). The best-fit gaussian minima remain close to the energy minima of the single molecule $E(\phi,\psi)$ maps, and are therefore significantly shifted apart from the conformations observed in the crystalline forms. It is interesting to note that the most important angular shifts do correspond to the most important shifts observed in the NMR spectra for amorphous and crystal states, respectively. As an example, $\phi_g=-10°$ for amorphous sucrose compared to a $\phi_c=-44.8°$ in the crystal, and it corresponds to an important shift in the CPMAS spectrum for the $C'_2$ fructose carbon. On the other hand, still for sucrose, the angle $\psi$ seems less affected ($\psi_g=100°$, $\psi_c=107.8°$) by amorphization, and correlatively, no important shift is evidenced in the NMR spectrum assigned to the glucose $C_1$ carbon. Similar angular and NMR shifts are also observed for lactose, but the poor quality of the best-fit obtained for the $C'_4$ glucose carbon prevents any conclusive quantitative results for this sugar.

This observation confirms the hypothesis made on the basis of the scrutiny of the experimental spectra. In other words, the average molecular conformation of disaccharides in glassy state is closer to a single molecule property than to that of the corresponding crystalline form, where long-range interactions prevail. This rather trivial observation should however be counterbalanced by noticing that the single molecule image is valid as a first approach, but is obviously not sufficient to completely account for the actual experimental CPMAS spectra of the glasses. While the $\sigma(C_x,\phi,\psi)$ surface can confidently be assumed to be an intramolecular property, this is not true anymore for the $\chi(\phi,\psi)$ conformational occupation probability density, that can be influenced by long range interactions. As a consequence, a suitably "fitted" gaussian $\chi(\phi,\psi)$ gives better results than the probability function provided by force-field based intramolecular $E(\phi,\psi)$.

Another result of our simulations is the relevant difference between the relative width of the $\chi(\phi,\psi)$ functions for trehalose and the other sugars. As reported in Table 4, and illustrated in Figure 4, the standard deviation of the probability curves for $\phi$ and $\psi$ torsional angles is as large as 50° for trehalose, and about 20° for lactose and sucrose. A similar conclusion has also been reached by using the Boltzmann probability function based on the calculated $E(\phi,\psi)$ surface, for which a quite unrealistic high temperature of 1800 K had to be adopted for trehalose (Table 4). Therefore, both the gaussian and the Boltzmann fits suggest that the conformational space explored by trehalose molecules in the amorphous phase should be larger than that calculated for the other two sugars, as well as for the single trehalose molecule. At a second order level, as already discussed in the previous paper [12, 18] and mentioned in another study on model trehalose molecule [12, 18] the unrealistic temperature value that has to be assumed suggests that a pure single molecule energy landscape model is not correct in case of amorphous trehalose, and that border-conformations in the higher energy regions could be equally made accessible.

On the basis of the convergent evidence, this result should be considered as model-independent, and mainly of intramolecular origin. Indeed, it is easily recognised that intramolecular hydrogen bonds are present in the crystal forms as well in solution of both lactose and sucrose [32-34, 37, 38], while these interactions are mostly absent in α,α-trehalose [32-34, 37, 38]. Our work therefore suggests that these intramolecular hydrogen bonds persist in glassy lactose and sucrose, stabilizing the molecular conformations close to the most probable one, while the intramolecular interactions that stabilize the single trehalose molecule may be easily balanced by the surrounding interactions providing comparatively enhanced flexibility around its glycosidic bond. Higher flexibility of trehalose in comparison with maltose and sucrose has also been deduced by Lerbret *et al.* [39] on the basis of molecular dynamics simulations of water solutions of these disaccharides.

To further explore and possibly corroborate the above reported idea, resort is made to an implicit concept that is beyond the influence of the surrounding random molecules on the intramolecular potential. Indeed, Scopigno *et al.* [40] found a correlation between the fragility of a glass-forming liquid, *m*, and the temperature coefficient of the non-ergodicity factor, *f*, given in the relation:

$$f^{-1}(Q \to 0, T) = 1 + \alpha \frac{T}{T_g} \qquad (4)$$

This factor, in the low T limit, is related to the vibrational properties of the glassy dynamics. In other words, Scopigno *et al.* claim that properties deriving from the curvature of the potential energy landscape are related to properties deriving from the viscosity increase upon super cooling [41]. They have found experimentally that the parameter α is proportional to the fragility m usually defined as

$$m = \frac{\partial(\log(\tau))}{\partial\left(\frac{T_g}{T}\right)}\Bigg|_{\frac{T_g}{T}=1} \tag{5}$$

Moreover, in a recent paper [41], Bordat *et al.* have reproduced numerically the proportionality between α and m and have demonstrated that a large fragility is correlated to the increase of anharmonicity and capacity for intermolecular coupling of the potential describing the interactions within the system considered.

Whether this correlation is a very general property of all materials or confined to "flexible" and "sticky" molecules like H-bond rich polyols is to be clarified. In our context, however, this parallel implies that, at least for homologous molecules, the change in the Δφ-Δψ values points to both the fragility and the non-ergodicity factor. The exploitation of this point is given in Figure 5, where the evolution of the inverse non-ergodicity parameter $f^{-1}(Q \to 0, T)$ obtained from MD simulations of glassy trehalose and sucrose is plotted as a function of the temperature scaled to the glass transition temperature, $T_g$ (see ref [42] for details).

The two curves of Figure 5 can be fitted according to equation (4) in the very low temperature domain, well below $T_g$, to get α=0.3582 and 0.1718 for trehalose and sucrose respectively.

Therefore, we can state that the larger fragility of the trehalose glass compared to the sucrose glass is a signature of enhanced intermolecular interactions. These enhanced intermolecular interactions take place in anharmonic potentials favouring then the flexibility of trehalose. So, a similar conclusion is reached on the flexibility of trehalose and its intermolecular coupling with surrounding molecules on the basis of different independent approaches, experimental NMR and simulations data, MD simulation [23, 39] and fragility analysis [41]. At that stage, it should be very interesting to compare the results of the present study to models that take into account the cooperativity in such glassy hydrogen-bonding networks at a longer length scale. In that sense, the tandem bilayer model proposed by Phillips [10] for trehalose is one of the possible sketches that are compatible with our experimental and simulation data.

Some other comments appear also necessary as far as it concerns the bioprotectant role of trehalose, that is the main reason for the converging interest in the properties of this molecule. Even not going to the several hypotheses made along the last years, the central question lies in the evolution of the intermolecular interactions that trehalose is able to develop upon increasing concentration from the dilute or semi-dilute solution to the eventually formed solid state surrounding the biosystems to be protected. All findings from dilute and semi-dilute solutions support a privileged interaction between sugar (trehalose in particular) and water, by displacing the "water-like structure" in favor of a "sugar-solvation structure" [26, 39]. Average conformational fluctuations around glycosidic dihedral angles for different solution composition at 200 K show little variance within the energy basin leading Sokolov et al [42] to argue little influence of intramolecular hydrogen bondings on the dynamics of the system. This work provides also a reinterpretation of the "folding" of trehalose onto a biomolecule and substituting water molecules during drying by rearranging its conformation [43]. However, the presence of at least one water molecule strongly bound to trehalose molecule in solution and in hydrated mixtures seems univocally assessed by several experimental and computational investigations. Although it has been previously suggested a possible pathway in the mechanism by which water hydration is substituted by sugar-sugar interaction [22], unfortunately, the consequence of this behaviour in the formation of the anhydrous sugar phase has not been fully exploited.

On the molecular level, fast rotational motions and higher $\nu OH$ at $T_g$ have been taken as indication that trehalose glass is less densely packed than sucrose glass, although these molecules have the same molecular weight. This fact lead the authors [44] to conclude that the molecular volume of trehalose is expanded either because the weaker intermolecular interactions or the higher molecular flexibility. The results of the present work do not provide a definitive answer to the way trehalose bioprotectant action is manifested but rather address again the attention on the subtle chameleon properties of trehalose adaptation. Work is in progress in our laboratories on inelastic scattering in the UV range [45] and on the possible effect of structural heterogeneity on the time scale of molecular transformations.

## Acknowledgments

A.C. is grateful to University Lille I for Visiting Professorship and to hosting LDSMM for hospitality.

*proportional to the fragility m can be obtained from the inverse non-ergodicity parameter f$^{-1}$ (determined from the total intermediate scattering functions) relative to T/T$_g$, well below T$_g$, in the glassy state. This method is entirely accessible by Molecular Dynamics simulations, which are shorter (sufficiently long to determine the non-ergodicity parameter f due to vibrations only). The 10 temperatures investigated for the three sugars are 25, 50, 100, 150, 175, 200, 225, 250, 300, 350 K. The technical details of simulations and the force field used are given elsewhere in Ref. [P. Bordat et al., Europhys. Lett.].*

**Table Captions**

Table 1 : Dihedral angles $\phi$ and $\psi$ defining the glycosidic torsion angles of $\alpha,\alpha$-trehalose, $\alpha$-lactose and sucrose.

Table 2 : Coefficients of the linear conversion $\sigma = A.S_{calc.} + B$ from calculated magnetic shielding $S_{calc}$ to chemical shift $\sigma$.

Table 3 : Lowest energy conformations ($\phi_0, \psi_0$) calculated from the BIO85 force field for trehalose, lactose and sucrose, compared to the dihedral angles ($\phi_c, \psi_c$) observed in T$\beta$ anhydrous trehalose [32], lactose monohydrate [33] and sucrose [34] crystalline forms.

Table 4 : Comparison of the best-fit populations $\chi(\phi,\psi)$ of either gaussian form (see Figure 4) or of Boltzmann repartition on the E($\phi,\psi$) maps calculated with the BIO85 force field (see Figure 2).

|   | α,α-trehalose | α-lactose | sucrose |
|---|---|---|---|
| φ | $O_5$-$C_1$-$O_1$-$C'_1$ | $O_5$-$C_1$-$O_1$-$C'_4$ | $O_5$-$C_1$-$O_1$-$C'_2$ |
| ψ | $C_1$-$O_1$-$C'_1$-$O'_5$ | $C_1$-$O_1$-$C'_4$-$C'_3$ | $C_1$-$O_1$-$C'_2$-$O'_2$ |

**Table 1**

|   | α,α-trehalose | α-lactose | sucrose |
|---|---|---|---|
| A | -1.4 | -1.1 | -1.02 |
| B | 240.1 | 216.3 | 201.34 |

**Table 2**

|   | α,α-trehalose | α-lactose | sucrose |
|---|---|---|---|
| Calculated ($\phi_0, \psi_0$) | (65°, 65°) | (-60°, 100°) | (-20°, 100°) |
| Crystal ($\phi_c, \psi_c$) | (60.1°, 60.8°) | (-92.6°, 94.6°) | (-44.8°, 107.8°) |

**Table 3**

|  | α,α-trehalose | α-lactose | sucrose |
|---|---|---|---|
| $(\phi_g, \psi_g)$ | (70°, 70°) | (-65°, 125°) | (-10°, 100°) |
| $\Delta\phi = \Delta\psi$ | 50° | 20° | 15° |
| $T_B$ | 1800 K | 300 K | 300 K |

**Table 4**

**Figure Captions**

Figure 1 : $^{13}$C CPMAS spectra of crystalline (a) and amorphous (b) disaccharides α,α-trehalose, α-lactose and sucrose. The assignment of the NMR lines is presented for the two carbons involved in the glycosidic bond.

Figure 2 : Ramachandran adiabatic maps E(φ,ψ) of α,α-trehalose, α-lactose and sucrose, calculated with the molecular mechanics BIO85 force field. The origin of energies is chosen at the lowest minimum for each sugar, and major contours are displayed each 5 kcal/mol.

Figure 3 : $^{13}$C isotropic chemical shift σ(C$_x$,φ,ψ) surfaces of the C$_x$ carbons involved in the glycosidic bond of α,α-trehalose, α-lactose and sucrose. The maps were calculated by DFT (B3PW91/3-21+G**) on the same molecular conformations defining the energy maps E(φ,ψ). The contours are displayed in ppm.

Figure 4 : Lower line : best fits of the chemical shift distributions for the glycosidic carbons of α,α-trehalose, α-lactose and sucrose, using gaussian (solid black line) or Boltzmann (solid grey line) probability density of conformational occupation χ(φ,ψ) ; comparison with the experimental CPMAS spectrum (dashed line). Upper line : corresponding best-fit gaussian probability densities χ(φ,ψ).

Figure 5 : Inverse non-ergodicity parameter $f^{-1}(Q \to 0, T)$ relative to T/T$_g$. $f(Q \to 0, T)$ has been obtained by a Q-quadratic extrapolation of $f(Q,T)$ determined from the total $F(Q,t)$ intermediate scattering functions at Q=Q$_{max}$ where Q$_{max}$ is the **position of the maximum of the first peak of the structure factor S(Q). The glass transition temperatures used for rescaling abscissa are the experimental $T_g$ = 393 K and 341 K for trehalose and sucrose, respectively.**

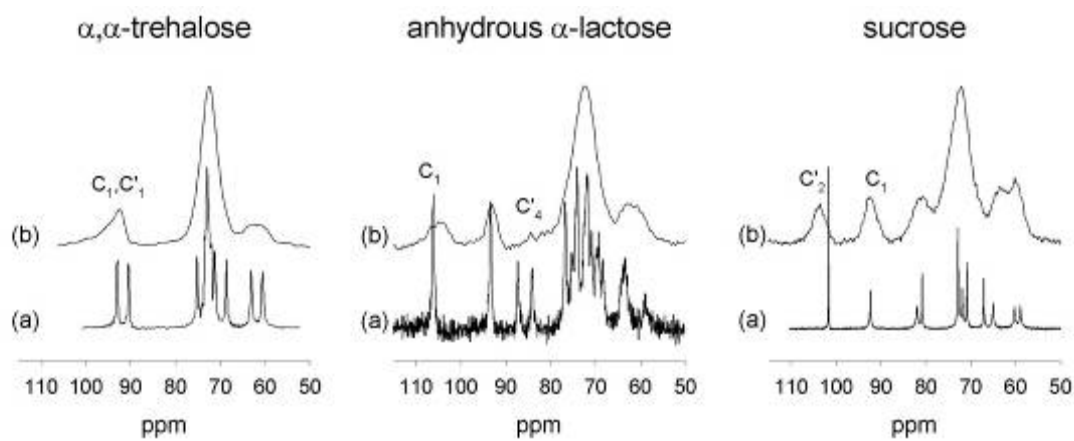

**Figure 1**

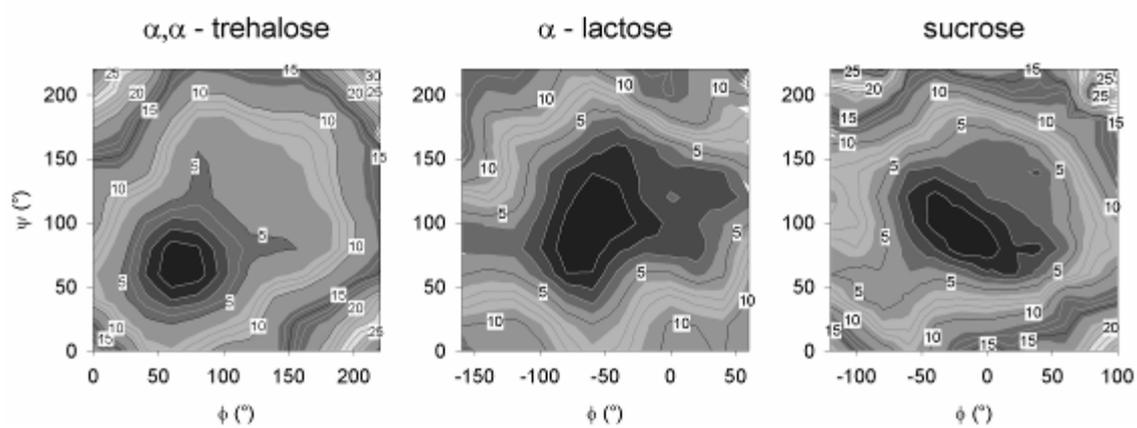

**Figure 2**

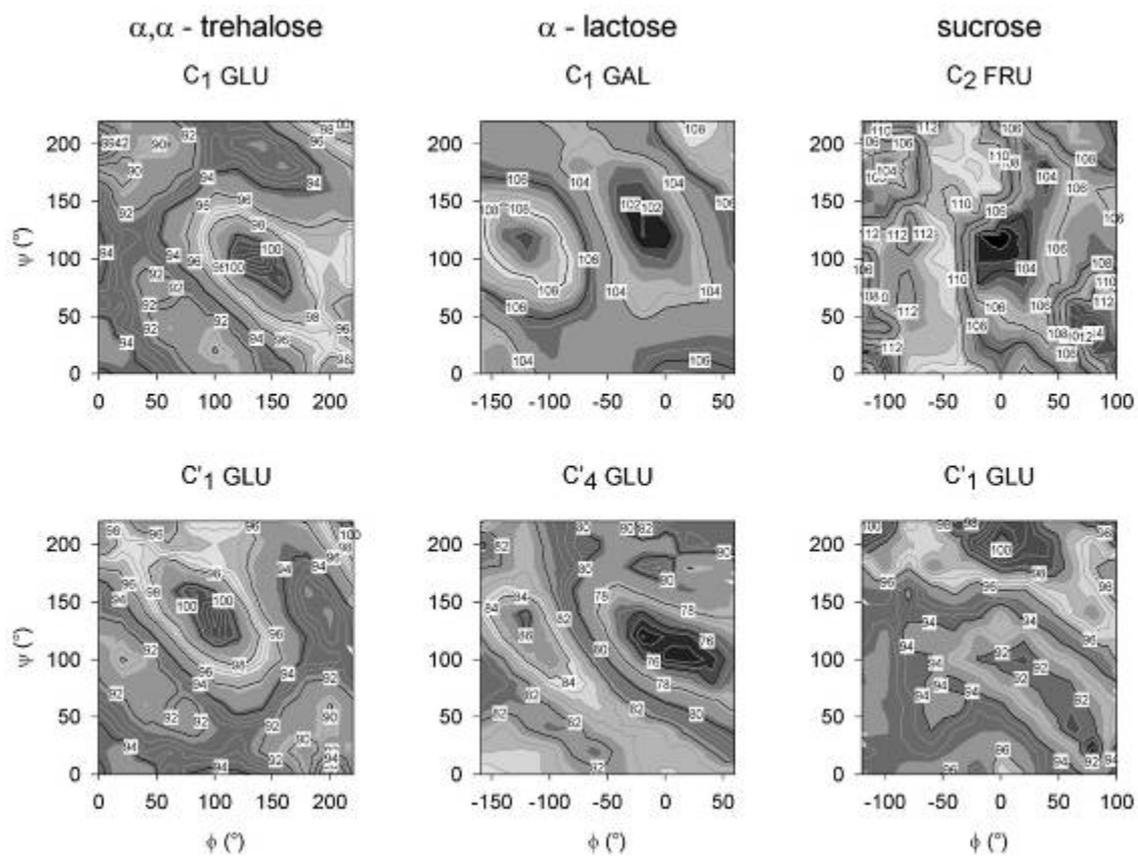

**Figure 3**

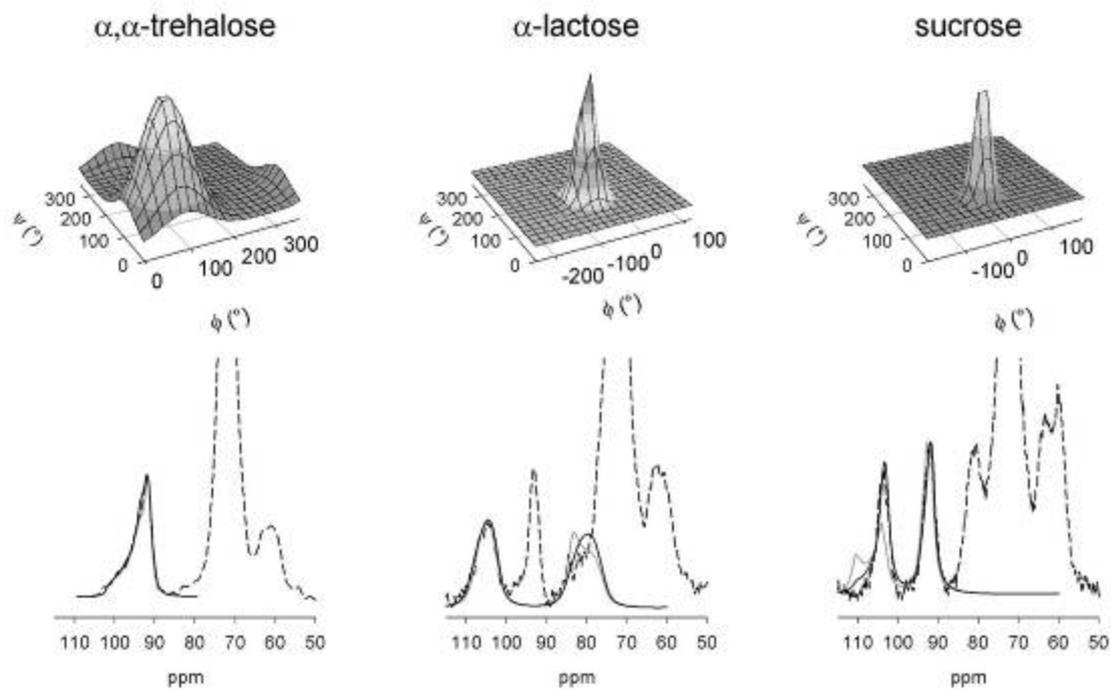

**Figure 4**

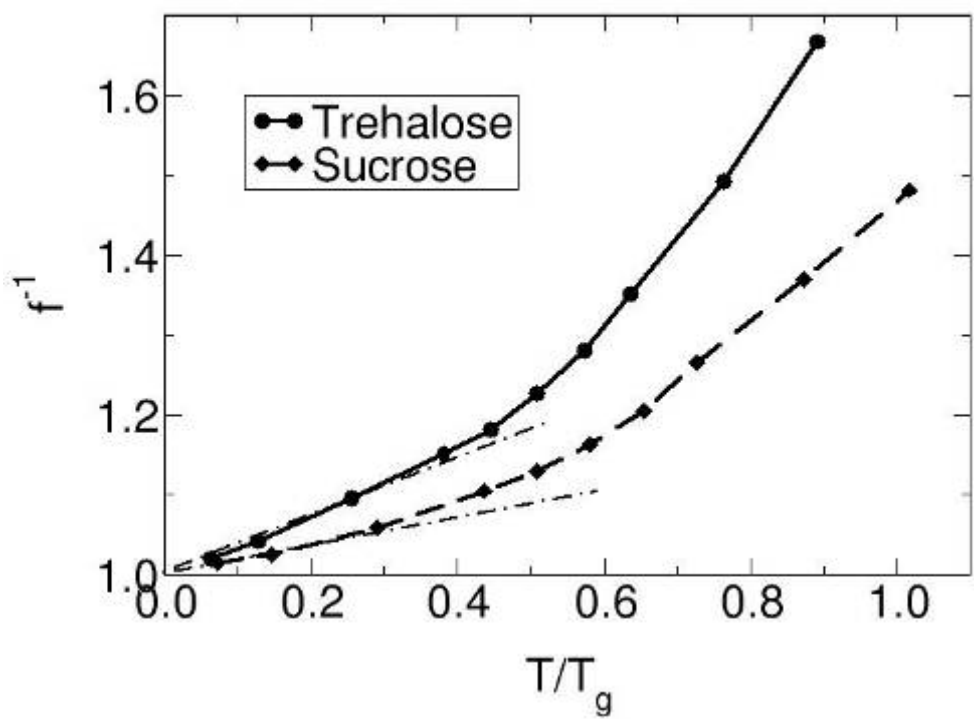

**Figure 5**